# Correlation between the Dzyaloshinskii-Moriya interaction and the orbital angular momentum at an oxide / ferromagnet interface

Hans T. Nembach[1,2], Emilie Jué[2,3], Eric R. Evarts[2#], Justin M. Shaw[2]

[1]JILA, University of Colorado, Boulder, CO 80309, USA

[2]Quantum Electromagnetics Division, National Institute of Standards and Technology, Boulder, CO 80305, USA

[3]Department of Physics, University of Colorado, Boulder, CO 80309, USA

We report on the Dzyaloshinskii-Moriya (DMI) interaction at the interface between a ferromagnet and an oxide. We demonstrate experimentally that oxides can give rise to DMI. By comparison of systems comprised of Pt/Co$_{90}$Fe$_{10}$/Oxide and Cu/Co$_{90}$Fe$_{10}$/Oxide, we also show how oxidation of one interface can enhance and add to the total DMI of that generated by the Pt interface. This is due to the fact that the DMI on both interfaces promotes left-handed chirality. Finally, by use of ferromagnetic resonance spectroscopy, we show that the DMI and the spectroscopic splitting factor, which is a measure of the orbital momentum, are correlated. This indicates the importance of hybridization and charge transfer at the oxide interface for the DMI.

*Introduction.* --The interface between heavy metals such as Pt and a ferromagnet can give rise to a variety of phenomena, where the strong spin-orbit coupling in the heavy metal plays an important role. These phenomena include perpendicular anisotropy, Dzyaloshinskii-Moriya interaction (DMI) and spin-orbit torques[1–6]. The DMI gives rise to a rich variety of chiral magnetic structures in thin magnetic films. For sufficiently large DMI, chiral spin-chains, chiral domain walls and skyrmions can be stabilized. Such spin textures are now being explored for novel electronic applications[7,8].

Fundamentally, the DMI is an anti-symmetric exchange interaction that requires a broken inversion symmetry in combination with spin-orbit interaction[1–3,9,10]. The interface formed between a ferromagnet and a heavy metal provides both the necessary symmetry breaking of the system and the large spin-orbit coupling. The DMI energy $E$ for two neighboring spins $\vec{S}_i$ and $\vec{S}_j$ at the interface is given by $E = -\vec{D}_{ij} \cdot (\vec{S}_i \times \vec{S}_j)$, where $\vec{D}_{ij}$ is the DMI vector. The relation between the DMI vector and the symmetry of the interface is given by the Moriya rules[9,11]. For highly symmetric interfaces such as (111)-oriented interfaces, the DMI vector $\vec{D}_{ij}$ is perpendicular to the plane of the triangle formed by $\vec{S}_i$ and $\vec{S}_j$ and a third atom in the heavy metal.

So far, not much is known about the underlying physics of DMI. Recently, a direct proportionality between the anti-symmetric and the symmetric Heisenberg exchange was shown in the Ni$_{80}$Fe$_{20}$/Pt system[12]. Such proportionality was theoretically predicted for bulk metallic spin-glasses and bulk magnetic oxides[9,10]. S. Kim et al.[13] have found that the DMI and the Heisenberg exchange have a similar temperature dependence in the Pt/Co/MgO system. Moreover, they used x-ray circular dichroism to determine the temperature dependence of the orbital momentum and established a correlation with the DMI. Relativistic first principles calculations by Belabbes et al. have shown that the sign and magnitude of the DMI in 3d-5d systems are correlated to the band filling determined by Hund's first rule[14]. Most experimental work on interfacial DMI has been confined to metallic materials, where the heavy metal is, for example, Pt or Ir[12]. It has also been found that the interface generated between graphene and a ferromagnet can exhibit a measurable DMI[15]. However, it is well known that *bulk* oxides can also have chiral spin-structures that originate from DMI. An example is Hematite, $\alpha$-$Fe_2O_3$, where the canted spin-structure gives rise to weak ferromagnetism[16,17]. Moriya found that the DMI is proportional to the symmetric superexchange *J* in magnetic oxides[9]: $|\vec{D}_{ij}| \sim \Delta g/g \cdot J$, where *g* the spectroscopic splitting factor and $\Delta g$ its deviation from the free electron value $g_e = 2$. The deviation $\Delta g = 2 \cdot l/s$ itself is proportional to the orbital momentum *l* of the electron with spin *s*, implying a relationship between the DMI and the orbital momentum.[1,9]

---

[#] Now at IBM Research, Albany, NY.

Recent density functional theory calculations (DFT) by Belabbes et al. for oxide/ferromagnet interfaces show a correlation of the DMI with an electric dipole moment at the interface[18]. This electrical dipole moment is the result of hybridization of the 3d-orbitals of the ferromagnet and the p-orbitals of the oxygen together with the associated charge transfer. The hybridization will also change the orbital angular momentum, which will alter the spectroscopic splitting factor. Thus, the interface between an oxide and a ferromagnet can give rise to DMI, while simultaneously modifying the orbital momentum at the interface. In this work, we show experimentally that the interface formed between an oxide and a metallic ferromagnet can give rise to DMI, and we gain deeper insight into the underlying physics that generates DMI through the comparison of the spectroscopic splitting factor *g* (related to the orbital momentum) to the magnitude of the DMI as the interface becomes increasingly oxidized. Here, we find a change of *g*, and thus a change in the orbital momentum, as the oxide forms. This change most likely originates from hybridization together with charge transfer at the interface. This implies a correlation of an electrical dipole moment at the interface with the DMI as predicted in previous DFT calculations.[18]

*Experiment*. -- We prepared two series of samples to experimentally determine the influence of oxide coverage of a ferromagnet on the DMI. One series had a Pt underlayer and a second, the control series, had a Cu underlayer. The Cu underlayer was chosen since it has small spin-orbit coupling compared to Pt. As a result, the presence of any DMI measured in the control samples would demonstrate that it originates at the oxide interface. The sample structure consists of substrate/Ta (3 nm)/Pt(6 nm)/$Co_{90}Fe_{10}$(2 nm)/Cu (3 nm)/Ta(2 nm) and substrate/Ta (3 nm)/Cu (6 nm)/$Co_{90}Fe_{10}$(2 nm)/ Cu (3 nm)/Ta(2 nm) and were sputter deposited on thermally oxidized Si substrates. Ta serves as an adhesion layer and induces a (111)-texture. Following deposition of the CoFe layer, we exposed the CoFe surface to gas mixture of 99% Ar and 1% Oxygen for a time *t* ranging from 0 s to 1000 s. This allowed us to systematically vary the amount of oxide formation. Following the oxidation process, the chamber was pumped for 5 minutes to a pressure better than 13 x $10^{-9}$ Pa (1 x $10^{-8}$ Torr) before continuing with the deposition of a Cu (3 nm)/Ta (2 nm) capping layer to prevent any additional oxidation.

We used x-ray diffraction to characterize structural changes of the samples during the oxidation process. The unoxidized Pt/$Co_{90}Fe_{10}$ sample has a smaller in-plane lattice constant of $0.3586 \pm 0.0004$ nm relative to Cu/$Co_{90}Fe_{10}$ which has a lattice constant of $0.3597 \pm 0.0002$ nm. The lattice constants for both sample series is continuously strained with increasing oxidation and after 1000 s of oxidation both sample series have within error bars the same in-plane lattice constant ($0.3604 \pm 0.0006$ nm for Pt/$Co_{90}Fe_{10}$ and $0.3606 \pm 0.0001$ nm for Cu/$Co_{90}Fe_{10}$). Thus, the oxide interface strains the Cu/$Co_{90}Fe_{10}$ lattice less than the Pt/$Co_{90}Fe_{10}$ lattice, see the SI for the x-ray data.

Measurements with a magnetometer based on a superconducting quantum interference device (SQUID) measurements show that the magnetic moment per area decreases with oxidation time. This trend is consistent with the thickness of the $Co_{90}Fe_{10}$ layer being reduced as a result of the formation of Co oxide and Fe oxide at the interface. The thickness reduction for the samples with the longest oxidation time of 1000 s is 13% and 12% for the Pt/$Co_{90}Fe_{10}$ and for the Cu/$Co_{90}Fe_{10}$ sample, respectively, or about 1.3 monolayer. This is consistent with the assumption that only one to two monolayers become oxidized. We do not expect the saturation magnetization $M_s$ itself would change significantly due to the strain induced during the oxidation process. This assumption is supported by the fact that $\mu_0 M_s$, where $\mu_0$ is the permeability of vacuum, for the unoxidized samples is $(1.898 \pm 0.001)$ T for the Pt/$Co_{90}Fe_{10}$ series and $(1.851 \pm 0.009)$ T for the Cu/$Co_{90}Fe_{10}$ series are within 2% of each other despite difference in their respective lattice constants. As a result, for a given sample series, we use the value of $M_s$ for the unoxidized sample for all calculations.

Broadband ferromagnetic resonance (FMR) was used in both the in-plane and in the out-of-plane geometries, where the magnetic field was swept through the resonance for a given frequency. The perpendicular anisotropy field $H_k$ and the spectroscopic splitting factors $g^{\parallel}$ and $g^{\perp}$ for the in-plane and out-of-plane geometry were determined from fits to the Kittel equations for the in-plane and out-of-plane geometries respectively:

$$f = \frac{\mu_0 \mu_B g^{\parallel}}{h} \sqrt{H \cdot (H + M_{eff})}$$

$$f = \frac{\mu_0 \mu_B g^{\perp}}{h} (H - M_{eff})$$

where $M_{eff} = M_s - H_k$ is the effective magnetization, $H$ is the applied magnetic field, $\mu_B$ is the Bohr magneton, $h$ the Planck constant.. We obtain $M_{eff}$ directly from fits to the Kittel equations, and further refine the values for $g^{\parallel}$ and $g^{\perp}$ using an asymptotic analysis[19]. In Fig. 2 (a) and 2(b), $g^{\perp}$ is shown for both samples series. $g^{\parallel}$ was not used in the correlation analysis since it has significantly larger error bars, which prohibits the determination of clear trend with the oxidation time. (See the SI for the data of $g^{\parallel}$). This is especially true for the Pt/$Co_{90}Fe_{10}$ series, where the spin-pumping significantly increases the linewidth and damping. The FMR data together with the SQUID magnetometry shows that the perpendicular anisotropy field $\mu_0 H_k$ is reduced during the oxidation process from $(1.110 \pm 0.003)$ T to $(0.73 \pm 0.01)$ T for the Pt/$Co_{90}Fe_{10}$ and from $(0.627 \pm 0.009)$ T to $(0.17 \pm 0.01)$ T for the Cu/$Co_{90}Fe_{10}$ series. After an oxidation time of 1000 s, the anisotropy of the oxide interface favors an in-plane orientation of the magnetization. Oxides are known to induce anisotropy. Easy-plane anisotropy has been found in the past for the CoFe/$CoFeO_x$ system[20], which is in agreement with findings in this study. These results are in contrast to interfaces of ferromagnets with MgO and $Al_2O_3$, which induce perpendicular anisotropy[5,21].

The DMI was determined using Brillouin-Light scattering spectroscopy (BLS) with a 532 nm wavelength laser. The photons are inelastically scattered by magnons, while energy and momentum are conserved.

The angle of incidence in our experiment is 45 degrees, which fixes the wavevector *k* of the magnons at 16.7 μm$^{-1}$. The frequency *f* for spin-waves propagating perpendicular to the applied magnetic field, the so-called Damon-Eshbach modes, in the presence of DMI is given by[22]:

$$f = \frac{\mu_0 \mu_B g^{\parallel}}{h} \sqrt{\left(H + \frac{2A}{\mu_0 M_s}k^2 - H_k + M_s \frac{1-e^{-t|k|}}{t|k|}\right) \cdot \left(H + \frac{2A}{\mu_0 M_s}k^2 + M_s\left(1 - \frac{1-e^{-t|k|}}{t|k|}\right)\right)} + sgn(M)\frac{\mu_B g^{\parallel}}{h} \cdot \frac{2D_{DMI}}{M_s}k$$

where *t* is the thickness of the ferromagnetic layer, *A* is the exchange constant and $D_{DMI}$ the volumetric DMI constant. $D_{DMI}$ gives the effective strength of the DMI, as if it would be a volume effect and not interfacial. The last term in this equation is dependent on both the spin-wave propagation direction (sign of the wavevector $k$) and field polarity. By measuring the spin-wave frequency for positive and negative field polarity with the BLS, we obtain the frequency-shift $\Delta f_{DMI} = (f(+H) - f(-H))/2$ due to the presence of the DMI. This frequency shift is equal to the last term in Eq. 3 and is shown in Fig. 1, where we plot the BLS spectra for the unoxidized sample of the Pt/Co$_{90}$Fe$_{10}$ series for both field polarities. The spin-wave frequencies for positive and negative fields are clearly distinguishable and shifted from one another due to the DMI at the Pt/Co$_{90}$Fe$_{10}$ interface. The error bars are based on one standard deviation.

It is also important to consider that Damon-Eshbach spin-waves have a non-uniform profile through the thickness of the ferromagnet with the largest amplitude at the top or the bottom interface depending on their propagation direction[23,24]. Consequently, in the presence of interfacial anisotropy, Damon-Eshbach spin-waves propagating in opposite directions can have different frequencies due to their non-reciprocal character. In order to estimate the frequency-shift caused by the change of the interfacial perpendicular anisotropy during the oxidation, we used the equation derived by O. Gladii et al.[25]:

$$\Delta f_{ani} \simeq \frac{8\gamma}{\pi^3}\frac{\Delta K}{M_s} \cdot \frac{k}{1 + \frac{\Lambda^2 \pi^2}{t^2}}$$

where $\Lambda = \sqrt{2A/\mu_0 M_s^2}$ is the exchange length, $\Delta K$ is the difference of the anisotropy of the top and bottom interfaces induced by the oxidation (see the SI for details), and *A* is the exchange constant. We calculated the oxide induced $\Delta f_{ani}$ using the changes of the perpendicular anisotropy measured by FMR and the spin-wave stiffness $D_{stiff} = 7.4 \cdot 10^{-40}$ Jm$^2$ from Grimsditch et al[26], with $A = D_{stiff} M_s/2g^{\parallel}\mu_B$. $D_{stiff}$ is not an interfacial property and we do not expect any change with oxidation. The largest anisotropy-induced frequency shift is $\Delta f_{ani} \simeq 10$MHz, which is small compared to the total oxide induced frequency-shift of about 100 MHz. In Fig. 2 we plot the DMI, which is calculated from the frequency shift after correcting for the anisotropy-induced shift, for both sample series. We used $g^{\parallel}$ determined by FMR for each sample and the saturation magnetization for the respective unoxidized sample from each series for the calculation. For both sample series the DMI is increasing with the oxidation time. The presence of a significant DMI for the control series after oxidation demonstrates that a ferromagnet/oxide interface can give rise to DMI. The DMI for the unoxidized Pt/Co$_{90}$Fe$_{10}$ sample originates from the DMI at the Pt interface. The oxidation further increases the total DMI indicating that the DMI of the oxide interface promotes the same left-handed chirality as the DMI at the Pt interface. We are using the convention that positive DMI gives rise to clockwise rotation of the spins (right-handed chirality), see SI. The total increase of the frequency-shift is larger for the Pt/Co$_{90}$Fe$_{10}$ series than for Cu/Co$_{90}$Fe$_{10}$ series. This increase can (within error bars) be accounted for by the reduction in thickness of the ferromagnet from the formation of the oxide. This increases the fractional contribution of the interfacial DMI at the Pt interface to the volume averaged DMI by ~0.1 mJ/m$^2$.

*Discussion.* – In Fig.2 (a) and (b), $g^\perp$ and $D_{DMI}$ both show a monotonic dependence with increasing oxidation time. $g^\perp$ has both a volume and an interfacial contribution, which would require a Co$_{90}$Fe$_{10}$ thickness series to deconvolute these two contributions. However, since the film thickness is on the order of the exchange length, a clear distinction between bulk and interfacial contributions to $g^\perp$ is not possible and we will not attempt to separate the two contributions. $D_{DMI}$, on the other hand, originates from the interface. Here, we compare the as measured values for $g^\perp$ and $D_{DMI}$ in order to avoid any ambiguity by correcting only $D_{DMI}$ for the thickness reduction during the oxidation. In the SI we also show a similar analysis, where instead we normalized $D_{DMI}$ to a constant thickness of 1 nm Co$_{90}$Fe$_{10}$. The overall trend between the two methods of thickness normalization remains unchanged. Only the quantitative change of $D_{DMI}$ becomes dependent on the method used. Furthermore, we quantify the correlation between $D_{DMI}$ and $g^\perp$ through the Pearson correlation coefficients, which assesses the linear relationship between two parameters. The Pearson correlation coefficient for $D_{DMI}$ and $g^\perp$ is $r_s^{Pt} = -0.88$ for the Pt/Co$_{90}$Fe$_{10}$/Oxide and $r_s^{Cu} = -0.98$ for the Cu/Co$_{90}$Fe$_{10}$/Oxide series. These values indicate that $D_{DMI}$ and $g^\perp$ are indeed correlated. This can be visualized in the plots of $D_{DMI}$ vs. $g^\perp$ as shown in the insets of Figs. 2 (a) and 2(b). Again, we do not pursue a correlation analysis of $g^\parallel$ due to the larger scatter of the data.

Density functional theory calculations (DFT) for the Ir/Fe/Oxide system show that the presence of oxide at the interface gives rise to DMI and is correlated with an electric dipole moment at the interface[18]. This electric dipole moment originates from hybridization and the associated charge transfer at the interface between the oxygen and metal atoms. This finding is in agreement with DFT calculations by H. Yang et al., which showed that the DMI for a Pt/Co/MgO trilayer is enhanced compared to a Pt/Co bilayer[27]. Moreover, they found that application of an electric field perpendicular to the surface of the trilayer structure further increases the DMI, when the positive electric field points from the insulator to the metal. The electric field gives rise to a charge redistribution in a manner similar to a material with high electronegativity. Oxygen, for example , has a Pauling electronegativity of 3.44, much higher than the electronegativity of elemental 3d ferromagnets, which have an electronegativity of about 1.9[28].

During the oxidation the 3d-orbitals of the Co and the Fe, which point towards the interface (d$_{xz}$, d$_{yz}$ and $d_{z^2}$), will hybridize with the oxygen p$_z$ orbitals. The d$_{xy}$ and d$_{x2-y2}$ orbitals are oriented within the interface plane, and as such, hybridize to a much lesser degree with the oxygen orbitals. The bonding between Co and O was confirmed by photoelectron x-ray spectroscopy (XPS) for the Co/AlO$_x$ systems[29]. The hybridization shifts the energy levels of the 3d-orbitals and changes their respective occupation. This also changes the associated total orbital momentum. Thus, the change of $g^\perp$, which originates from a change in the orbital momentum *l*, is indirect evidence of the hybridization and the associated charge transfer at the oxide interface. This charge transfer, which depends on the difference of the electronegativity of the materials on both sides of the interface, increases the asymmetry of the electronic structure at the interface. Such lack of inversion symmetry is a critical requirement for DMI.

The hybridization at the oxide interface is also related to the change of the interfacial anisotropy. The respective band levels depend on the orientation of the magnetization, which gives rise to an orientation dependence of the orbital momentum. The difference of the orbital momentum for the out-of-plane plane and in-plane geometry $\Delta l^{\perp-\parallel} = l^{\perp} - l^{\parallel}$ is, according to Bruno's theory, directly related to the interfacial perpendicular anisotropy and was extensively studied for many systems experimentally and theoretically[5,30–32]. The orbital asymmetry $\Delta l^{\perp-\parallel}$ is proportional to the difference of the out-of-plane and in plane g-factors. In Fig. 3 we plot the perpendicular anisotropy field $H_k$ vs $g^{\perp} - g^{\parallel}$ for the Cu/Co$_{90}$Fe$_{10}$/Oxide series and find qualitative agreement with Bruno's model. However, the large scatter and the larger error bars for $g^{\parallel}$ found in the data for the Pt/Co$_{90}$Fe$_{10}$/Oxide series obscures any relationship between $H_k$ and $g^{\perp} - g^{\parallel}$.

The theory by Moriya predicts a correlation of $D_{DMI}$ with $g$. A direct comparison between Moriya's theory and our result is difficult. First, Moriya's theory was developed for ionic bonds between oxygen and Fe in the bulk, where the spins are interacting by superexchange and $g$ is considered isotropic. Here, the DMI is interfacial and the bonds are characterized by hybridization between the Co and Fe d-orbitals and the oxygen p-orbitals. Thus, it is not necessarily expected that the DMI shows the same trend with $g$ as Moriya's theory predicts. Nevertheless, we still find that $g^{\perp}$ and $D_{DMI}$ are correlated as originally predicted. It is important to keep in mind that there are other contributing factors that determine the magnitude of the DMI. For example, the strength of the Heisenberg exchange can also affect the magnitude of the DMI. Therefore, all of these factors must be taken into consideration.

In summary, we have demonstrated that ferromagnetic / oxide interfaces can induce DMI. In metal/Co$_{90}$Fe$_{10}$/oxide systems give rise to left-handed chiral spin chains (negative DMI value). Thus, it will lead to an enhancement of the total DMI in a Pt/ Co$_{90}$Fe$_{10}$/Oxide trilayer system. Moreover, $g^{\perp}$ follows the same trend as the DMI with the oxidation time. This is in agreement with DFT calculations by Belabbes et al. for the Fe/Oxide system, which show an increase of hybridization at the interface together with charge transfer gives rise to DMI. The electric dipole moment, which result from the charge transfer, is correlated to the DMI. Finally, we show that the orbital asymmetry for the Cu/Co$_{90}$Fe$_{10}$/Oxide series is correlated with the anisotropy.

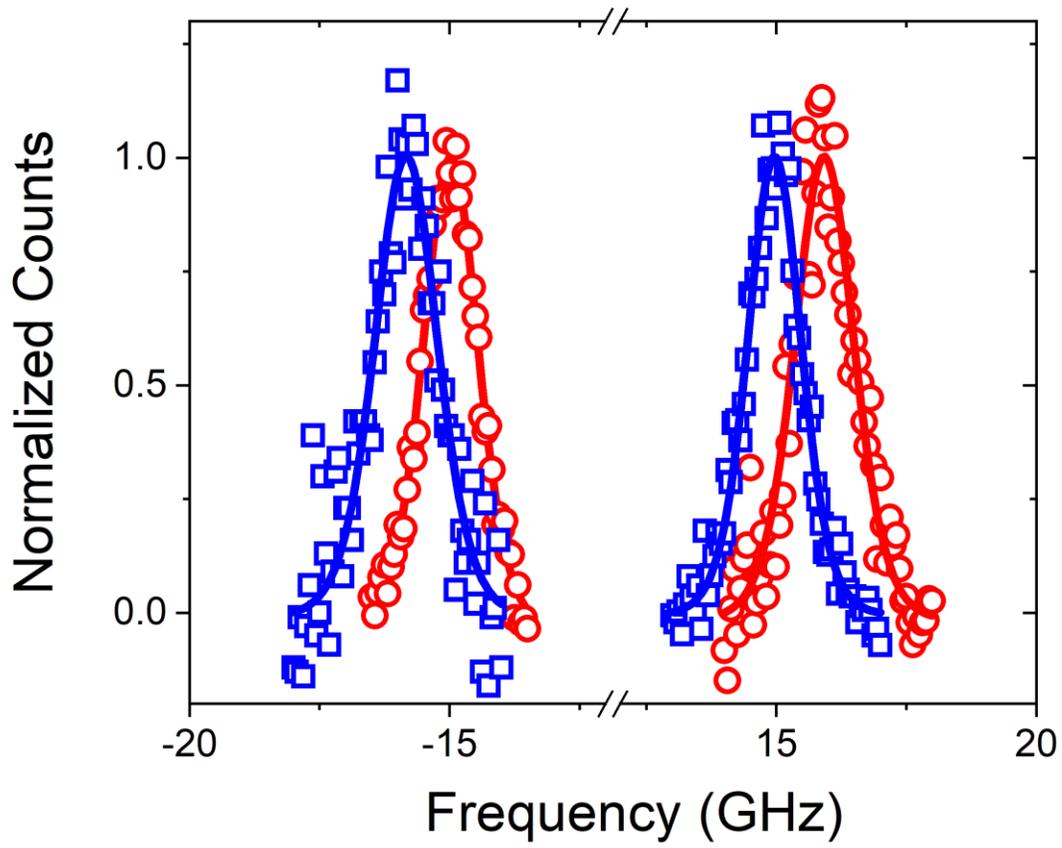

Fig. 1: BLS spectra for the two field polarities for the unoxidized Pt/Co$_{90}$Fe$_{10}$ sample positive field (blue) and negative field (red).

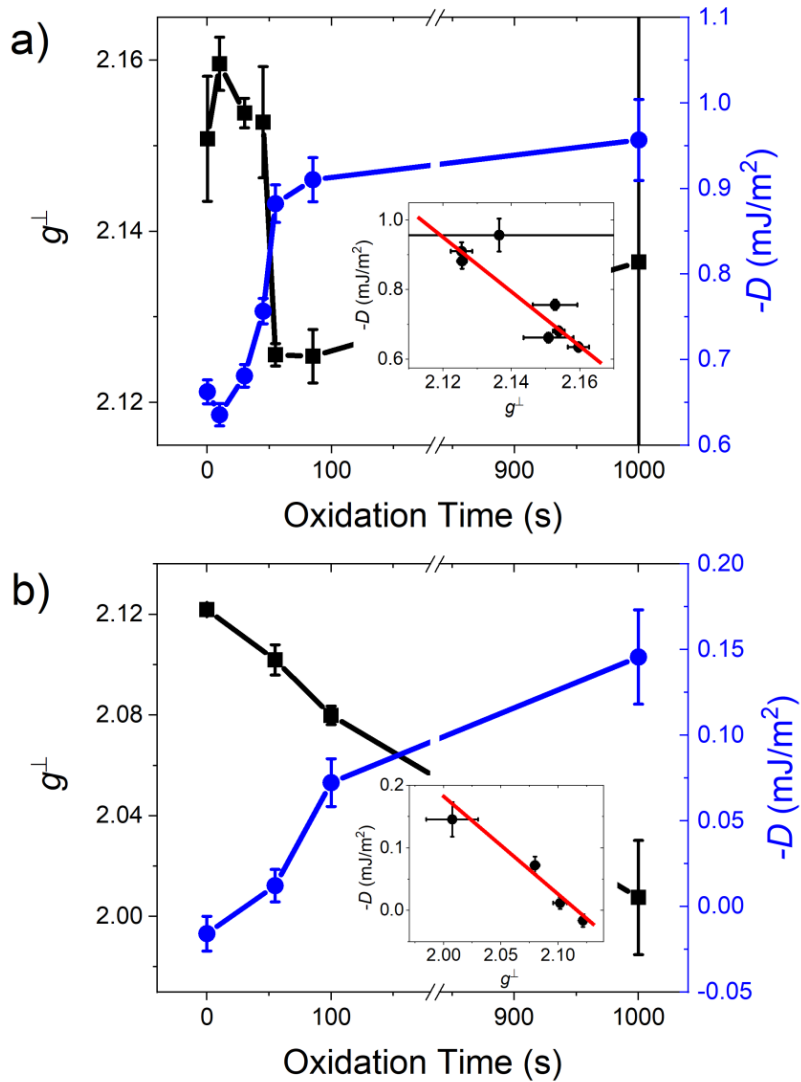

Fig.2: The change of the magnitude of the DMI is shown in a) for the Pt/Co$_{90}$Fe$_{10}$ and in b) for the Cu/Co$_{90}$Fe$_{10}$ sample series with increasing oxidation time (blue circles). The spectroscopic splitting factor $g^\perp$ is represented by the black squares in the two respective graphs. $g^\perp$ shows the opposite trend than $D$ with oxidation time. The insets in a) and b) show $D$ vs. $g^\perp$ for the two sample series. The insets demonstrate the correlation between these two quantities. The red line is a guide to the eye.

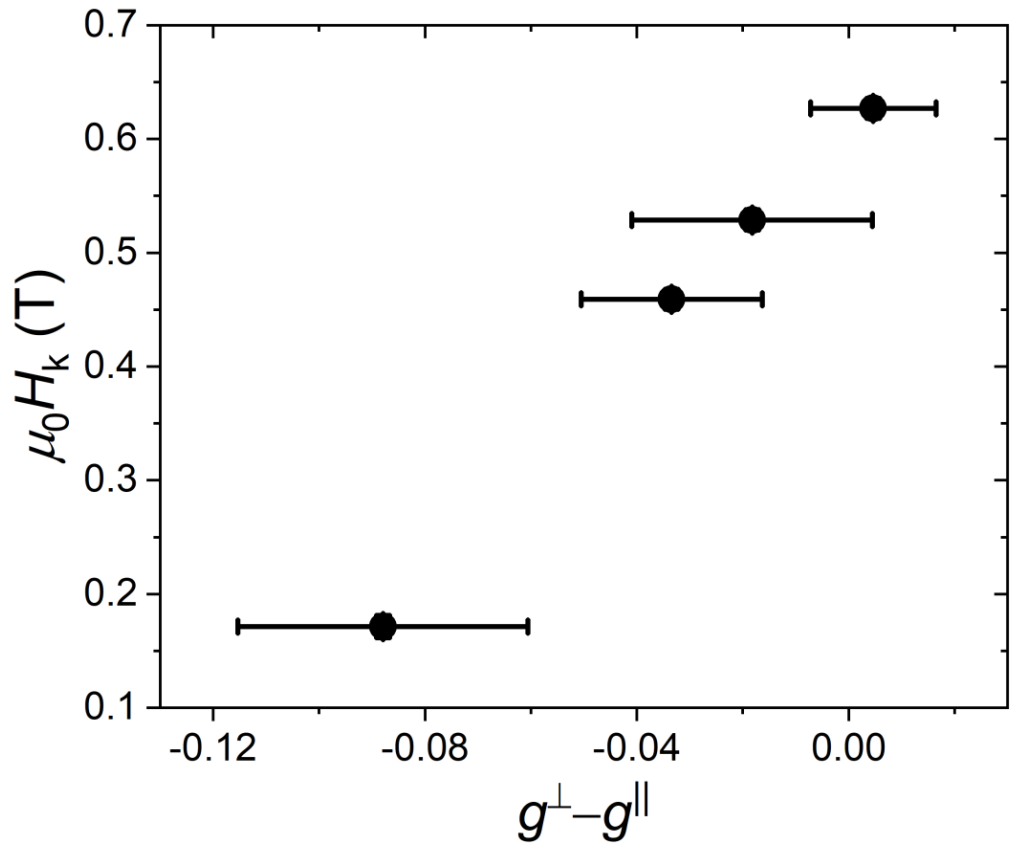

Fig. 3: Perpendicular anisotropy vs. the difference between the spectroscopic g-factor for the out-of-plane geometry and the in-plane geometry, which is proportional to the difference of the orbital momentum for these two geometries, for the $Cu/Co_{90}Fe_{10}$ sample series.

1. Moriya, T. New Mechanism of Anisotropic Superexchange Interaction. *Phys. Rev. Lett.* **4**, 228–230 (1960).

2. Dzyaloshinsky, I. A thermodynamic theory of "weak" ferromagnetism of antiferromagnetics. *J. Phys. Chem. Solids* **4**, 241–255 (1958).

3. Fert, A. Magnetic and transport Properties of metallic multilayers. *Mater. Sci. Forum* **59&60**, 439 (1990).

4. Manchon, A. *et al.* Current-induced spin-orbit torques in ferromagnetic and antiferromagnetic systems. (2018).

5. Dieny, B. & Chshiev, M. Perpendicular magnetic anisotropy at transition metal/oxide interfaces and applications. *Rev. Mod. Phys.* **89**, 025008 (2017).

6. Hellman, F. *et al.* Interface-induced phenomena in magnetism. *Rev. Mod. Phys.* **89**, 025006 (2017).

7. Fert, A., Cros, V. & Sampaio, J. Skyrmions on the track. *Nat. Nanotechnol.* **8**, 152–156 (2013).

8. Kang, W., Huang, Y., Zhang, X., Zhou, Y. & Zhao, W. Skyrmion-Electronics: An Overview and Outlook. *Proc. IEEE* **104**, 2040–2061 (2016).

9. Moriya, T. Anisotropic Superexchange Interaction and Weak Ferromagnetism. *Phys. Rev.* **120**, 91–98 (1960).

10. Fert, A. & Levy, P. M. Role of Anisotropic Exchange Interactions in Determining the Properties of Spin-Glasses. *Phys. Rev. Lett.* **44**, 1538–1541 (1980).

11. Hoffmann, M. *et al.* Antiskyrmions stabilized at interfaces by anisotropic Dzyaloshinskii-Moriya interaction. *Nat. Commun.* **8**, (2017).

12. Nembach, H. T., Shaw, J. M., Weiler, M., Jué, E. & Silva, T. J. Linear relation between Heisenberg exchange and interfacial Dzyaloshinskii-Moriya interaction in metal films. *Nat. Phys.* **11**, 825–829 (2015).

13. Kim, S. *et al.* Correlation of the Dzyaloshinskii–Moriya interaction with Heisenberg exchange and orbital asphericity. *Nat. Commun.* **9**, 1648 (2018).

14. Belabbes, A., Bihlmayer, G., Bechstedt, F., Blügel, S. & Manchon, A. Hund's Rule-Driven Dzyaloshinskii-Moriya Interaction at 3 d − 5 d Interfaces. *Phys. Rev. Lett.* **117**, (2016).

15. Yang, H. *et al.* Significant Dzyaloshinskii–Moriya interaction at graphene–ferromagnet interfaces due to the Rashba effect. *Nat. Mater.* **17**, 605 (2018).

16. Aharoni, A., Frei, E. H. & Schieber, M. Curie point and origin of weak ferromagnetism in hematite. *Phys. Rev.* **127**, 439 (1962).

17. Sandratskii, L. M. & Kübler, J. First-principles LSDF study of weak ferromagnetism in Fe2O3. *EPL Europhys. Lett.* **33**, 447 (1996).

18. Belabbes, A., Bihlmayer, G., Blügel, S. & Manchon, A. Oxygen-enabled control of Dzyaloshinskii-Moriya Interaction in ultra-thin magnetic films. *Sci. Rep.* **6**, (2016).

19. Shaw, J. M., Nembach, H. T., Silva, T. J. & Boone, C. T. Precise determination of the spectroscopic g-factor by use of broadband ferromagnetic resonance spectroscopya). *J. Appl. Phys.* **114**, 243906 (2013).

20. Monso, S. *et al.* Crossover from in-plane to perpendicular anisotropy in Pt/CoFe/AlOx sandwiches as a function of Al oxidation: A very accurate control of the oxidation of tunnel barriers. *Appl. Phys. Lett.* **80**, 4157–4159 (2002).

21. Ikeda, S. *et al.* A perpendicular-anisotropy CoFeB–MgO magnetic tunnel junction. *Nat. Mater.* **9**, 721–724 (2010).

22. Moon, J.-H. *et al.* Spin-wave propagation in the presence of interfacial Dzyaloshinskii-Moriya interaction. *Phys. Rev. B* **88**, 184404 (2013).

23. Damon, R. W. & Eshbach, J. R. Magnetostatic modes of a ferromagnet slab. *J. Phys. Chem. Solids* **19**, 308–320 (1961).

24. Hurben, M. J. & Patton, C. E. Theory of magnetostatic waves for in-plane magnetized anisotropic films. *J. Magn. Magn. Mater.* **163**, 39–69 (1996).

25. Gladii, O., Haidar, M., Henry, Y., Kostylev, M. & Bailleul, M. Frequency nonreciprocity of surface spin wave in permalloy thin films. *Phys. Rev. B* **93**, 054430 (2016).

26. Grimsditch, M., Fullerton, E. E. & Stamps, R. L. Exchange and anisotropy effects on spin waves in epitaxial Co films. *Phys. Rev. B* **56**, 2617–2622 (1997).

27. Yang, H., Boulle, O., Cros, V., Fert, A. & Chshiev, M. Controlling Dzyaloshinskii-Moriya Interaction via Chirality Dependent Atomic-Layer Stacking, Insulator Capping and Electric Field. *Sci. Rep.* **8**, 12356 (2018).

28. *CRC Handbook of Chemistry and Physics, 99th Edition*. (CRC Press, 2018).

29. Manchon, A. *et al.* Analysis of oxygen induced anisotropy crossover in Pt/Co/MOx trilayers. *J. Appl. Phys.* **104**, 043914 (2008).

30. Yang, H. X. *et al.* First-principles investigation of the very large perpendicular magnetic anisotropy at Fe$|$MgO and Co$|$MgO interfaces. *Phys. Rev. B* **84**, 054401 (2011).

31. Shaw, J. M., Nembach, H. T. & Silva, T. J. Measurement of orbital asymmetry and strain in Co 90 Fe 10 /Ni multilayers and alloys: Origins of perpendicular anisotropy. *Phys. Rev. B* **87**, (2013).

32. Bruno, P. Tight-binding approach to the orbital magnetic moment and magnetocrystalline anisotropy of transition-metal monolayers. *Phys. Rev. B* **39**, 865–868 (1989).

**Sign convention for the Dzyaloshinskii-Moriya Interaction**

The chirality of the magnetization texture is determined by the sign of DMI. We are using the convention that positive DMI gives rise to clockwise rotation of the spins corresponding to a right-handed chirality, Fig. S1. With this convention, two interfaces, which induce the same chirality (same DMI sign), increase the total DMI value of the multilayer stack, whereas two interfaces, that induce the opposite chirality (opposite DMI signs), reduce the total DMI value.

Clockwise rotation, D>0, right-handed chirality:

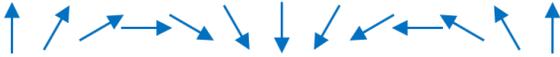

Counterclockwise rotation, D<0, left-handed chirality:

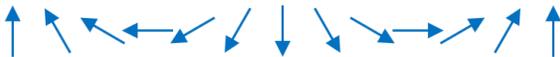

*Figure S1: Illustration of the convention used in this paper between the sign of DMI, the rotation of the spins and the chirality.*

**X-ray analysis of the crystal structure:**

We used x-ray diffraction to determine the change of the in-plane lattice constant during the oxidation process. The results are shown in Fig. S2. The oxidation continuously strains the lattice, which is more pronounced for the Pt/$Co_{90}Fe_{10}$ series than for the control series. After oxidation for 1000 s the in-plane lattice constant for both sample series are equal within error bars.

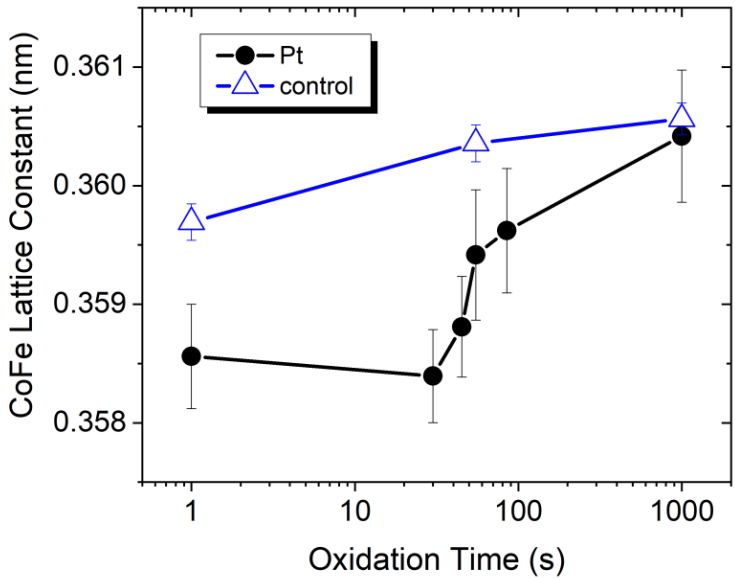

*Figure S2: In-plane lattice constant determined by x-ray diffraction for both sample series.*

**Determination of the difference of the interface anisotropy $\Delta K$ between the top and the bottom interfaces**

The calculation of the non-reciprocal frequency-shift $f_{NR}$ due to the non-uniform amplitude of the Damon-Eshbach spinwaves through the thickness of the ferromagnet requires the difference of the interface anisotropy between the top and bottom interface $\Delta K$, see eq. 4 in the main part. In the following, we will assume for simplicity that all anisotropy originates from the interface, implying that the volume anisotropy of the $Co_{90}Fe_{10}$ layer is zero. The unoxidized sample of the $Cu/Co_{90}Fe_{10}$ series has Cu on both interfaces of the ferromagnet. Under the assumption that both interfaces are identical, we can obtain the interfacial anisotropy $K_{Cu}$ for the Cu interface from $M_{eff}$ and $M_s$ measured by FMR and SQUID, respectively. This then allows us to determine the interface anisotropy for the Pt interface $K_{Pt}$ and the respective oxide interfaces.

**Dependence of the thickness $t$ of the ferromagnet on the oxidation**

We determined the thickness of the ferromagnet by measuring the saturation magnetization $M_s$ with SQUID magnetometry. The unoxidized samples are chosen as a reference. Any changes of the measured magnetic moment is attributed to a change in thickness. The results are shown in Fig. S3.

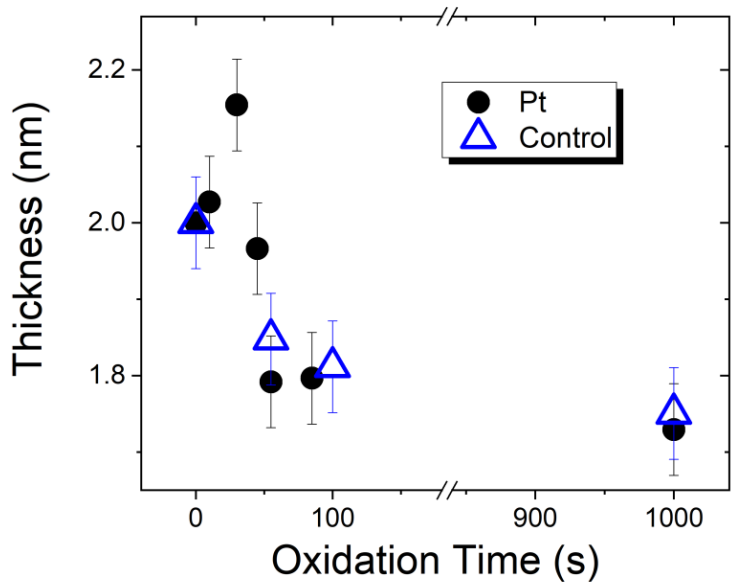

Figure S3: Reduction of the thickness of the ferromagnet with increasing oxidation time.

## $D$ normalized to 1 nm thickness of the ferromagnet

The DMI is an interface effect and as such the volume averaged value depends on the thickness for the ferromagnet. In order to account for the thickness changes, we normalize the DMI to a ferromagnet thickness of 1 nm. We use the thickness for the ferromagnet determined from SQUID measurements for the normalization, where we assume that any change in the magnetic moment during the oxidation is due to a change in the thickness. In Fig. S4 a) and b) the normalized values $D^{norm}$ for the Pt/Co$_{90}$Fe$_{10}$ and for the Cu/Co$_{90}$Fe$_{10}$ samples are shown together with $g^\perp$. It can be seen, that the correlation between the DMI and $g^\perp$ is maintained after normalization.

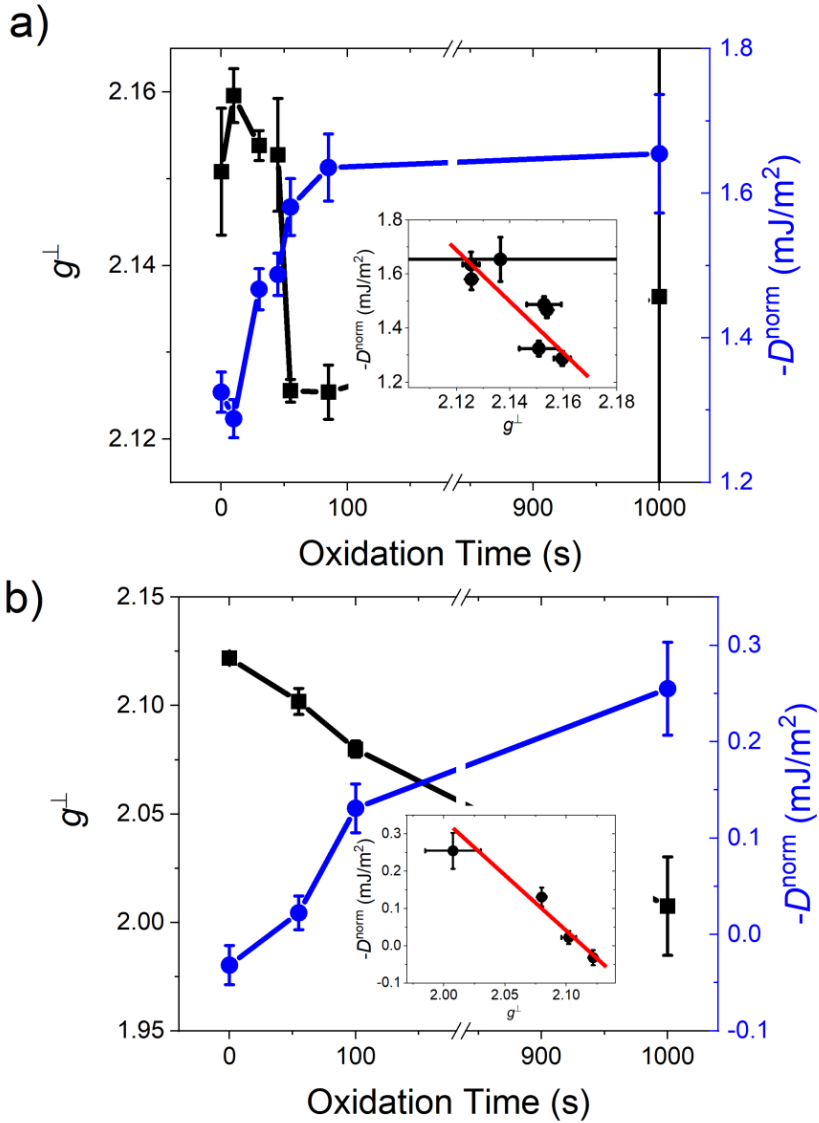

*Figure S4: $g^\perp$ and $D^{norm}$, which is the DMI normalized to a ferromagnet thickness of 1 nm for a) the Pt/Co$_{90}$Fe$_{10}$ and b) for the Cu/Co$_{90}$Fe$_{10}$ sample.*

### $H_k$ normalized to a 1 nm thickness of the ferromagnet

The anisotropy mainly originates from the two interfaces. In the same way we normalized the DMI shown in Fig. S4, we normalized $H_k$ for the Cu/Co$_{90}$Fe$_{10}$ samples series to a ferromagnet with a thickness of 1 nm. The proportionality between $H_k$ and the orbital asymmetry remains unchanged, see Fig. S5.

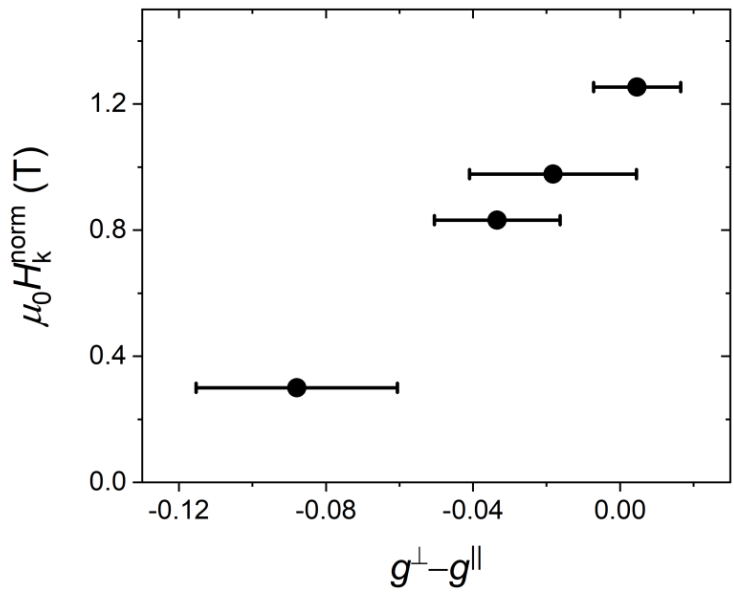

*Figure S5: The normalized perpendicular anisotropy $H_k^{norm}$ for the Cu/Co$_{90}$Fe$_{10}$ sample shows a linear dependence on the difference of the spectroscopic g-factor for the out-of-plane geometry and the in-plane geometry very similar to the not normalized value of $H_k$.*

## Dependence of $g^{\parallel}$ on the oxidation

We determined $g^{\parallel}$ for both sample series from FMR measurements. The results are shown in Fig. S6 and the error bars are significantly larger than for $g^{\perp}$.

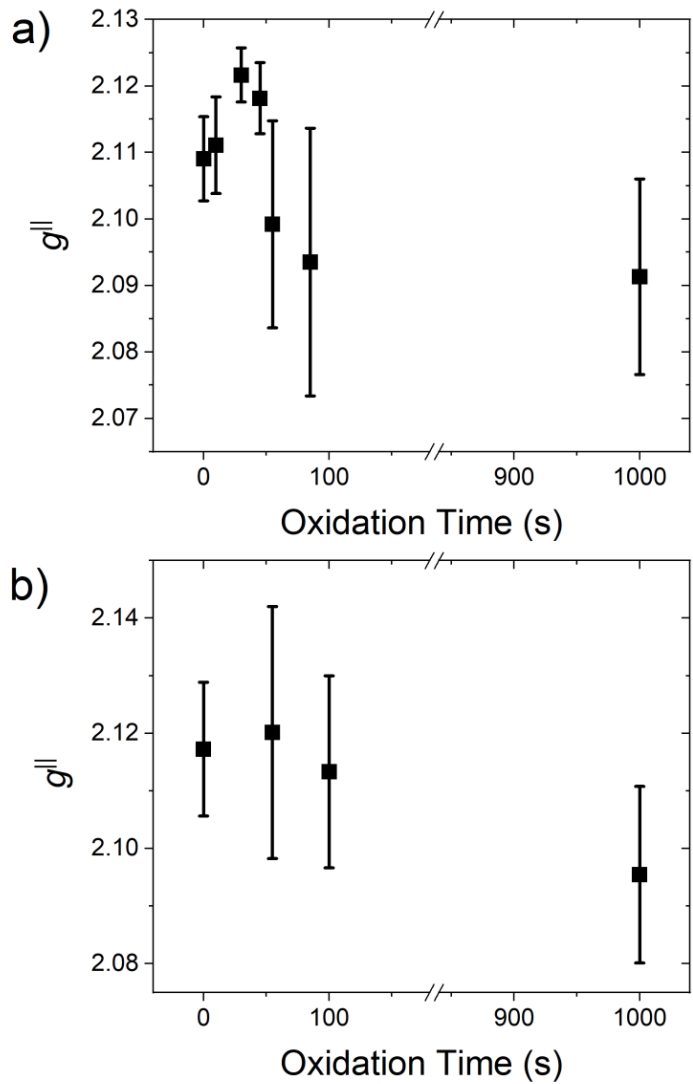

Figure S6: $g^{\parallel}$ for a) the Pt/Co$_{90}$Fe$_{10}$ and b) for the Cu/Co$_{90}$Fe$_{10}$ sample series.